\documentclass[a4paper,12pt, english]{article}
\usepackage[T1]{fontenc}
\usepackage{german}
\usepackage[latin1]{inputenc}
\usepackage{amsmath}
\usepackage{babel}
\usepackage{graphics}
\usepackage{amssymb}
\usepackage{amsfonts}
\usepackage{geometry}
\begin{document}

\title{Linear Response in the s.c. Bogolyubov model of Bose-Einstein Condensation.}

\author{L. A. Bányai, A. M. Bundaru\thanks{Permanent adress: Institute of Physics and Nuclear Engineering,
Bucharest } , and P. Gartner\thanks{Present address: Institut
 für  Theoretische Physik, Universität,  Bremen;Permanent address: National Institute for Materials Physics, Bucharest }
\\
Institut für Theoretische Physik, J.W. Goethe-Universität,
  Frankfurt am Main }
\date{}
\maketitle

\begin{abstract}
We discuss the linear response of the Bogolyubov theory of a weakly
interacting Bose gas below the critical temperature. The specific
aspects, due to induced perturbation terms in the self-consistent
treatment are discussed in detail. We show that Bogolyubov's model
having a gap-less equilibrium spectrum nevertheless gives rise to
an effective susceptibility which does not satisfy the super-fluidity
criterion of Hohenberg and Martin. 
\end{abstract}

\section{Introduction.}

The theoretical interest for Bose-Einstein Condensation (BEC) has
largely increased in the last years due to spectacular experimental
evidence with cold atoms in a magnetic trap \cite{Anderson,Davis}.
Most of the theoretical discussion for weakly interacting bosons is
going within the frame of the equilibrium Bogolyubov model \cite{Bogolyubov}.
(See some recent books \cite{Pethick,Pitaevskii} for the recent status
of the theory as well as a discussion of this model related to some
rigorous results in \cite{Zagrebnov}). This model is actually also
a self-consistent one, being however a truncation of the general self-consistent
(s.c.) Hartree-Fock (HF) approximation with the purpose to obtain
a gap-less quasi-particle spectrum, the former being from the variational
point of view thermodynamically better. On the other hand, the evolution
in real time towards equilibrium and the response of condensed bosonic
systems to time-dependent external perturbations require a consideration
of the self-consistent scheme out of equilibrium. Outside equilibrium
the self-consistent averages, including the anomalous ones like the
order parameter, are time-dependent and give rise to an induced perturbation.
This aspect known also in the ordinary theory Hartree-Fock theory
of Coulomb interacting particles modifies essentialy the theory of
linear response of a weakly interacting Bose gas. We revisit here
the linear response properties of the Bogolyubov theory . Our analysis
shows, that the Martin- Hohenberg \cite{Hohenberg} criterion for
supra-fluidity is actually not satisfied in this model. Although the
density-density fluctuation at $k=0$ diverges logarithmically at
small frequencies \cite{Pitaevskii}, the effective susceptibility
remains finite.

\section{The self-consistent Hartree-Fock approximation for weakly interacting
bosons.}

As it is well-known, the macro-canonical density matrix

\[
\rho=\frac{1}{Z}e^{-\beta(H-\mu N)}\]
 is the solution of the following variational problem for the free
energy:

\[
G=inf_{\rho}Tr\left\{ \rho\left(H-\mu N+\frac{1}{\beta}\ln(\rho)\right)\right\} \]
 with $\rho\geq0$ and $Tr\rho=1$ .

The equilibrium Hartree -Fock approximation is the approximate solution
of this problem with the variational reservoir \[
\rho_{test}=\frac{1}{Z}e^{-\beta(H_{test}-\mu N)}\]
 where $H_{test}$ is the most general one-particle test Hamiltonian.

It is known \cite{Cloizeaux}, that if

\[
H_{test}-\mu N=\sum_{\alpha}c_{\alpha}A_{\alpha}\]
 with $A_{\alpha}$ given operators and $c_{\alpha}$ the corresponding
variational parameters, the minimum of the above variational problem
is achieved for

\[
c_{\alpha}=\frac{\partial\langle H-\mu N\rangle}{\partial\langle A_{\alpha}\rangle}\qquad\text{where}\qquad\langle\cdots\rangle\equiv Tr\left(\rho_{test}\cdots\right)\,.\]
 If the operators $A_{\alpha}$ are bilinear and linear combinations
of creation and annihilation operators, then the (first) Wick theorem
for equilibrium averages can be used to express $<H-\mu N>$ as a
function of $<A_{\alpha}>$ for all $\alpha$ 's. This gives rise
to the usual HF coefficients. While applying these results to bosonic
systems one should take into account the existence of anomalous averages.

Actually the s.c. Hartree-Fock Hamiltonian is determined by the variational
principle up to a constant ({}``HF constant'') , which may be fixed
by the requirement

\[
\langle H\rangle=\langle H_{test}\rangle\,.\]

Here of course the above described Wick 's theorem must be applied
also on the left hand side of this equation.

The resulting s.c. Hartree-Fock Hamiltonian may be generalized successfully
under the same form also to non-equilibrium, however all the non-equilibrium
the averages are now time dependent. It may be shown, that within
this scheme (in the absence of time dependent external potentials),
although the HF Hamiltonian is time-dependent, the average HF energy
is still conserved.

Let us consider now the Hamiltonian of interacting bosons (in the
presence of an external potential $U(x)$ )

\begin{equation}
H=\int dx\left\{ \psi(x)^{+}\left(-\frac{\hbar^{2}}{2m}\nabla^{2}+U(x)\right)\psi(x)+\frac{1}{2}\int dx'w(x-x')\psi(x)^{+}\psi(x')^{+}\psi(x')\psi(x)\right\} \,.\label{hamiltonian}\end{equation}

This Hamiltonian conserves the number of bosons, but we are looking
for a spontaneous symmetry breaking solution. Therefore in applying
the HF-scheme we choose for the set $A_{\alpha}$ the operators $\psi(x)^{+}\psi(x'),\quad\psi(x)\psi(x'),\quad\psi(x)$
and their conjugates.

It can be shown that, for the case of spontaneous symmetry breaking
with anomalous averages $\langle\psi(x\rangle\rangle$ and $\langle\psi(x')\psi(x)\rangle$,
the first Wick theorem needed by the above stated recipe for equilibrium
gives rise to

\begin{eqnarray*}
\langle\psi(x)^{+}\psi(x')^{+}\psi(x')\psi(x)\rangle & = & \langle\psi(x)^{+}\psi(x')\rangle\langle\psi(x')^{+}\psi(x)\rangle+\langle\psi(x)^{+}\psi(x)\rangle\langle\psi(x')^{+}\psi(x')\rangle\\
 & + & \langle\psi(x)^{+}\psi(x')^{+}\rangle\langle\psi(x')\psi(x)\rangle-2|\langle\psi(x)\rangle\langle\psi(x')\rangle|^{2}\,.\end{eqnarray*}
 The last term is the correction to the usual Wick theorem. This result
may be obtained also by separating $\psi(x)=\psi'(x)+\langle\psi(x)\rangle$
and afterwards applying the usual Wick theorem. After reintroducing
the old operators one gets the previous result.

Based on this Wick's theorem one obtains immediately, according to
the recipe described above, the form of the s.c. HF Hamiltonian $H_{HF}^{0}$
(the upper index $0$ indicates, that it refers to equilibrium)

\begin{eqnarray}
H_{HF}^{0} & = & \int dx\left\{ \psi(x)^{+}\left(-\frac{\hbar^{2}}{2m}\nabla^{2}+U(x)\right)\psi(x)\right.\nonumber \\
 & + & \frac{1}{2}\int dx'w(x-x')\left[2\psi(x)^{+}\psi(x')\langle\psi(x)^{+}\psi(x')\rangle^{*}\right.\nonumber \\
 & + & 2\psi(x)^{+}\psi(x)\langle\psi(x')^{+}\psi(x')\rangle\nonumber \\
 & + & \left(\psi(x)\psi(x')\langle\psi(x)\psi(x')\rangle^{*}+h.c.\right)\nonumber \\
 & - & \left.\left.\left(4\psi(x)\langle\psi(x)\rangle^{*}|\langle\psi(x')\rangle|^{2}+h.c.\right)\right]\right\} \,.\label{HF-ham}\end{eqnarray}
 Here we took into account also that $w(x-x')=w(x'-x)$ and omitted
the HF constant.

It is important to remark, that the above described procedure differs
from a formal shift directly in the Hamiltonian followed by applying
the operatorial (second) Wick theorem for contractions.

We shall not discuss here furthermore the existence and subtle definition
of the Bose condensation in the presence of an external potential
(see \cite{condvol}), but we shall treat later the linear response
to a time dependent external potential.

The above described s.c. Hartree-Fock Hamiltonian may be generalized
also to non-equilibrium situations and in the case of a contact interaction
is often called as The Bogolyubov-Popov model\cite{Popov,Griffin}.

In the absence of an external potential $U(x)$ we shall rewrite the
s.c. HF Hamiltonian in the $k$-space. One introduces the Fourier
transforms

\[
\psi(x)=\frac{1}{\sqrt{V}}\sum_{k}b_{k}e^{\imath kx};\: w(x)=\frac{1}{V}\sum_{k}w_{k}e^{\imath kx}\]
 and one gets due to translational invariance (momentum conservation)

\begin{eqnarray}
H_{HF}^{0} & = & \sum_{k}\left\{ \left[\frac{\hbar^{2}k^{2}}{2m}+\frac{1}{V}\sum_{p}(w_{k+p}+w_{0})\langle b_{p}^{+}b_{p}\rangle\right]b_{k}^{+}b_{k}\right)\nonumber \\
 & + & \left.\frac{1}{2V}\sum_{p}w_{k+p}\left[\langle b_{p}b_{-p}\rangle^{*}b_{k}b_{-k}+h.c\right]\right\} \nonumber \\
 & - & \frac{2}{V}w_{0}|\langle b_{0}\rangle|^{2}\left(\langle b_{0}\rangle^{*}b_{0}+h.c.\right)\,,\label{hamHF01}\end{eqnarray}
 or

\begin{equation}
H_{HF}^{0}=\sum_{k}\{ e_{k}b_{k}^{+}b_{k}+\frac{1}{2}(c_{k}^{*}b_{k}b_{-k}+c_{k}b_{-k}^{+}b_{k}^{+})\}-2w_{0}\frac{1}{V}|\langle b_{0}\rangle|^{2}(\langle b_{0}\rangle^{*}b_{0}+\langle b_{0}\rangle b_{0}^{+})\label{hamHF02}\end{equation}
 with the notations

\begin{equation}
e_{k}\equiv\frac{\hbar^{2}k^{2}}{2m}+\frac{1}{V}\sum_{p}(w_{k+p}+w_{0})\langle b_{p}^{+}b_{p}\rangle;\:\label{ek}\end{equation}

\begin{equation}
c_{k}\equiv\frac{1}{V}\sum_{p}w_{k+p}\langle b_{p}b_{-p}\rangle\label{ck}\end{equation}

\section{The s.c. equilibrium HF solution for interacting bosons.}

In the case of a homogenous system in equilibrium one may study in
more detail the self-consistency equations and therefore the properties
of the solution. With the help of the Bogolyubov transformation one
may bring the Hamiltonian Eq.(\ref{hamHF02}) to the diagonal form

\[
H_{HF}^{0}=\sum_{k}\epsilon_{k}\alpha_{k}^{+}\alpha_{k}+const.\]
with some new (bosonic) quasiparticle annihilation and creation operators
$\alpha_{k},\alpha_{k}^{+}$.

\textbf{1)} Keeping only the terms of the HF Hamiltonian containing
operators of the non-condensate, relevant for the elementary excitations,
we have:\[
\frac{1}{2}\sum_{k\neq0}\left\{ (e_{k}-\mu)(b_{k}^{+}b_{k}+b_{-k}^{+}b_{-k})+c_{k}^{*}b_{k}b_{-k}+c_{k}b_{-k}^{+}b_{k}^{+}\right\} \]
 with $\mu$ being the chemical potential.

This Hamiltonian is diagonalized to the form $\sum_{k\neq0}\epsilon_{k}\alpha_{k}^{+}\alpha_{k}$
(up to a constant term) by the canonical transformation \cite{Bogolyubov}
to the new boson fields $\alpha_{k}$ and $\alpha_{-k}$ : \begin{equation}
b_{k}=u_{k}\alpha_{k}+v_{k}\alpha_{-k}^{+}\label{canon-k}\end{equation}

with \[
u_{k}=u_{-k};\quad v_{k}=v_{-k}\quad\text{and}\quad|u_{k}|^{2}-|v_{k}|^{2}=1\,.\]

One gets

\begin{equation}
u_{k}=\cosh(\chi_{k})e^{\imath\phi_{k}};\: v_{k}=-\sinh(\chi_{k})e^{-\imath\phi_{k}}e^{-\imath2\psi_{k}}\end{equation}

with $\psi_{k}$ defined by \begin{equation}
c_{k}=|c_{k}|e^{-\imath2\psi_{k}}\end{equation}

and

\begin{equation}
\tanh(2\chi_{k})=\frac{|c_{k}|}{e_{k}-\mu}\,.\end{equation}

One has to require of course

\[
\frac{|c_{k}|}{|e_{k}-\mu|}<1\,.\]

The phase $\phi_{k}$ remains arbitrary and the energies are given
by\[
\epsilon_{k}=sign(e_{k}-\mu)\sqrt{(e_{k}-\mu)^{2}-|c_{k}|^{2}}\,.\]
 For reasons of stability one should have

\[
e_{k}-\mu>0\]
 and therefore

\begin{equation}
\epsilon_{k}=\sqrt{(e_{k}-\mu)^{2}-|c_{k}|^{2}}\,.\label{eps-k}\end{equation}

Both inequalities may be combined to a single one

\begin{equation}
e_{k}-\mu>|c_{k}|\,.\end{equation}

\textbf{2)} Now one must consider also the part of the Hamiltonian
containing only the operators of the condensate ($\vec{k}=0$)

\[
(e_{0}-\mu)b_{0}^{+}b_{0}+\frac{1}{2}c_{0}b_{0}^{+}b_{0}^{+}+\frac{1}{2}c_{0}^{*}b_{0}b_{0}-2w_{0}\sqrt{V}|P|^{2}(P^{*}b_{0}+Pb_{0}^{+})\,.\]

Here we introduced the notation \begin{equation}
P\equiv\frac{1}{\sqrt{V}}\langle b_{o}\rangle\,.\label{P}\end{equation}
 The diagonalization of the bilinear terms is achieved by a similar
canonical transformation

\begin{equation}
b_{0}=u_{0}\bar{\alpha}_{0}+v_{0}\bar{\alpha}_{0}^{+}\end{equation}
 with

\begin{equation}
\tanh2\chi_{0}=-\frac{|c_{0}|}{\mu-e_{0}}\end{equation}

and the excitation energy is

\begin{equation}
\epsilon_{0}=\sqrt{(e_{0}-\mu)^{2}-|c_{0}|^{2}}\end{equation}

having the similar condition \begin{equation}
e_{0}-\mu>|c_{0}|\,.\label{sign0}\end{equation}

But we have still to shift the operators $\bar{\alpha}_{0}$ in order
to eliminate the linear terms\[
-2w_{0}\sqrt{V}|P|^{2}(P^{*}u_{0}+Pv_{0}^{*})\bar{\alpha}_{0}+h.c.\]

Then with

\begin{equation}
\alpha_{0}\equiv\bar{\alpha}_{0}-\frac{2w_{0}\sqrt{V}|P|^{2}}{\epsilon_{0}}(Pu_{0}^{*}+P^{*}v_{0})\end{equation}

we get a complete diagonalization in the form $\epsilon_{0}\alpha_{0}^{+}\alpha_{0}$
(up to a constant term). \\

\textbf{3)} Now we must proceed to discuss the self-consistency requirements
stemming from the fact, that the system is in macro-canonical equilibrium.
Then

\begin{equation}
\langle\alpha_{k}\rangle=0\quad\text{and}\quad\langle\alpha_{k}^{+}\alpha_{k}\rangle=\frac{1}{e^{\beta\epsilon_{k}}-1}\,.\end{equation}

Since $\langle b_{0}\rangle=u_{0}\langle\bar{\alpha}_{0}\rangle+v_{0}\langle\bar{\alpha}_{0}\rangle^{*}$
it follows

\begin{equation}
P=-\frac{2w_{0}}{\epsilon_{0}^{2}}|P|^{2}\left((\mu-e_{0})P+c_{0}P^{*}\right)\label{sc3}\end{equation}

to be a self-consistency equation.

Since the phase of $P$ may be chosen arbitrarily by a simple phase
transformation of the operators $b_{0}$ , it is convenient to chose
$P$ to be real and positive. Eq.(\ref{sc3}) may have a symmetry
breaking solution ($P\neq0$) satisfying

\begin{equation}
\frac{2w_{0}}{\epsilon_{0}^{2}}(e_{0}-\mu-c_{0})P^{2}=1\label{symbreak}\end{equation}
 with any real $c_{0}$, since for the positivity of $\epsilon_{0}$
we had to require already Eq.(\ref{sign0}).

For $\epsilon_{0}>0$, $\frac{1}{V}\langle\hat{\alpha}_{0}^{+}\hat{\alpha}_{0}\rangle\to0$
for $V\to\infty$. Then it follows also in the thermodynamic
limit \begin{equation}
\frac{1}{V}\langle b_{0}^{+}b_{0}\rangle=|P|^{2}\qquad\text{and}\qquad\frac{1}{V}\langle b_{0}b_{0}\rangle=P^{2}\,.\label{P2}\end{equation}

The non-condensate parameters in equilibrium are given by

\begin{equation}
f_{k}\equiv\langle b_{k}^{+}b_{k}\rangle=\frac{e_{k}-\mu}{\epsilon_{k}}\frac{1}{(e^{\beta\epsilon_{k}}-1)}+\frac{1}{2}(\frac{e_{k}-\mu}{\epsilon_{k}}-1)\end{equation}

and \begin{equation}
F_{k}\equiv\langle b_{k}b_{-k}\rangle=-\frac{c_{k}}{\epsilon_{k}}\left(\frac{1}{e^{\beta\epsilon_{k}}-1}+\frac{1}{2}\right)\,.\end{equation}

Thus the phase $-2\psi_{k}$ of $c_{k}$ differs by $\pi$ from the
phase of $F_{k}$.

We may go now overall to the thermodynamic limit after separating
the $\vec{p}=0$ terms. We have in the thermodynamic limit the functions
$w(k),\: f(k),\: F(k),\: c(k),\: e(k)$ and $\epsilon(k)$ and therefore
using the definition of $c_{k}$ Eq.(\ref{ck}) we get

\[
c(k)\equiv\frac{1}{(2\pi)^{3}}\int d^{3}pw(k+p)F(p)+w(k)P^{2}\]

or

\begin{equation}
c(k)\equiv-\frac{1}{(2\pi)^{3}}\int d^{3}pw(k+p)\frac{c(p)}{\epsilon(p)}\left(\frac{1}{e^{\beta\epsilon(p)}-1}+\frac{1}{2}\right)+w(p)P^{2}\label{sc2}\end{equation}

with the excitation energies \begin{equation}
\epsilon(k)=\sqrt{(e(k)-\mu)^{2}-|c(k)|^{2}}\end{equation}

Then it follows from the last equation , due to the reality of $w(k)$
also, that $c(k)$ must be real (its imaginary part satisfies a homogeneous
equation).

The parameters of the condensate are determined by

\begin{equation}
c_{0}=-\frac{1}{(2\pi)^{3}}\int d^{3}pw(p)\frac{c(p)}{\epsilon(p)}\left(\frac{1}{e^{\beta\epsilon(p)}-1}+\frac{1}{2}\right)+w(0)P^{2}\end{equation}

and the relation Eq.(\ref{symbreak}).

The chemical potential may be found from the average particle number
by

\begin{equation}
n_{0}+\frac{1}{(2\pi)^{3}}\int d^{3}kf(k)=n\,.\label{sc1}\end{equation}
 (here $n_{0}=P^{2}$ ), or \emph{\begin{equation}
P^{2}+\frac{1}{(2\pi)^{3}}\int d^{3}p\left(\frac{e(p)-\mu}{\epsilon(p)}\frac{1}{(e^{\beta\epsilon(p)}-1)}+\frac{1}{2}(\frac{e(p)-\mu}{\epsilon(p)}-1)\right)=n\:.\label{sc1a}\end{equation}
} \\

In the case of a contact potential its Fourier transform is constant
$w(k)=w$ and as a consequence, also $c(k)=c$ with

\[
c=\frac{wP^{2}}{1+\frac{w}{(2\pi)^{3}}\int d^{3}p\frac{1}{\epsilon(p)}\left(\frac{1}{e^{\beta\epsilon(p)}-1}+\frac{1}{2}\right)}\]

However without a cut-off the integral diverges and therefore $c=0$\\

One should mention also, that for $T=0^{\circ}$ these HF results
coincide with those of the self-consistent variational model elaborated
by Girardeau and Arnowitt \cite{Girardeau}. \\

On the other hand, the Hugenholtz-Pines theorem \cite{Hugenholtz-Pines}
(which is a generalization of the Goldstone theorem \cite{Goldstone}
to non-vanishing temperatures ), requires that the spectrum of elementary
excitations at vanishing momentum starts with vanishing energy.

Unfortunately, within the s.c. HF scheme the theorem is not satisfied
and the elementary excitations start with a gap, except if $|c(0)|^{2}=|e(0)-\mu|^{2}$.
One may still try to chose a $w(q)$ to satisfy this condition \cite{Haug}. 

The results of this section are well-known in the literature (see
\cite{Haug,Huse} for variational approaches and other simplified
presentations\cite{Griffin} ). Here they are derived using des Cloizeaux's
elegant formalism for the sake of further reference.

\section{The Bogolyubov model.}

The Bogolyubov theory \cite{Bogolyubov} considers a contact interaction
between the bosons, but in contrast to the full s.c. HF theory, it
disregards the fluctuational part of the anomalous propagators. Therefore,
while obviously from the variational point of view it is poorer, it
obeys the Hugenholtz-Pines theorem. It may be regarded as the solution
of the variational problem with a constraint given by the known property
of the exact solution. We shall analyze in the next Section, which
are the consequences of this fact for the linear response from the
point of view of the super-fluidity condition of Martin and Hohenberg
\cite{Hohenberg}. In this Section we just describe the model itself.

Although the original theory of Bogolyubov \cite{Bogolyubov} was
formulated for a spacial homogeneous system, we shall define here
also its slight generalization in the presence of an external (time
dependent) potential .

In the Bogolyubov model one disregards the fluctuational part of the
anomalous propagators $\langle(\psi(x)-\langle\psi(x)\rangle)(\psi(x)-\langle\psi(x)\rangle)\rangle$
and considers the s.c. Hamiltonian

\begin{eqnarray}
{\mathcal{H}}_{B}(t) & = & \int dx\left\{ \psi(x)^{+}\left(-\frac{\hbar^{2}}{2m}\nabla^{2}+U(x,t)-\mu\right)\psi(x)\right.\nonumber \\
 & + & \frac{w}{2}\left[4\psi(x)^{+}\psi(x)\langle\psi(x)^{+}\psi(x)\rangle^{*}\right.\nonumber \\
 & + & \left(\psi^{+}(x)\psi^{+}(x)\langle\psi(x)\rangle^{2}+h.c.\right)\nonumber \\
 & - & \left.\left.\left(4\psi(x)\langle\psi(x)\rangle^{*}|\langle\psi(x)\rangle|^{2}+h.c.\right)\right]\right\} \,.\label{ham-HFB}\end{eqnarray}

Then one remains only with three s.c. entities $\langle\psi(x)^{+}\psi(x')\rangle,$
$\langle\psi(x)\rangle$ , $\langle\psi(x)\rangle^{*}$.

In the absence of any potential the equilibrium values of the averages
in the homogenous system emerge from those of the previous Section
after the replacement $\langle\psi(x)\psi(x')\rangle_{0}$ by $\langle\psi\rangle_{0}\langle\psi\rangle_{0}\equiv P^{2}$
Since, we treat here a contact potential we have a unique coefficient
$c$, which in the absence of the anomalous propagator is given by
\begin{equation}
c=w\frac{1}{V}\langle b_{o}\rangle^{2}=wP^{2}\label{cBog}\end{equation}
 and we choose again $P$ to be real and positive.

The equilibrium excitation spectrum is \begin{equation}
\epsilon_{k}=\sqrt{(e_{k}-\mu)^{2}-|c|^{2}}\,.\end{equation}
 where we took into account that $e(q)-\mu\geq0$ .

The analysis of the $k=0$ part of the Hamiltonian gives rise to the
s.c. condition for the existence of the anomalous solution Eq.(\ref{symbreak}),
which may be written as

\begin{equation}
e_{0}-\mu-c=\frac{\epsilon_{0}^{2}}{2wP^{2}}\,.\label{sc3a}\end{equation}
 From the last three equations results that $\epsilon_{0}=0$ and
therefore it follows \begin{equation}
e_{0}-\mu=c=wP^{2}\;.\label{emu}\end{equation}

To conclude, in the Bogolyubov model (in the absence of an external
potential) in the thermodynamic limit

\begin{equation}
\epsilon(q)=\sqrt{2wP^{2}\frac{\hbar^{2}}{2m}q^{2}+\left(\frac{\hbar^{2}q^{2}}{2m}\right)^{2}}\end{equation}

\[
u(q)=\sqrt{\frac{1}{2}\left(\frac{e(q)-\mu}{\epsilon(q)}+1\right)};\qquad v(q)=\sqrt{\frac{1}{2}\left(\frac{e(q)-\mu}{\epsilon(q)}-1\right)}\,,\]
 Hence the equilibrium excitation spectrum is gap-less and starts
linearly with the momentum.

The average particle number is given by

\[
P^{2}+\frac{1}{(2\pi)^{3}}\int d^{3}p\left(\frac{e(p)-\mu}{\epsilon(p)}\frac{1}{(e^{\beta\epsilon(p)}-1)}+\frac{1}{2}(\frac{e(p)-\mu}{\epsilon(p)}-1)\right)=n\,.\]

The introduction of a cut-off here is not necessary, since the integral
converges for the Bogolyubov spectrum, which is linear in $p$ at
small momenta and quadratic at large momenta.

\section{Linear response.}

Although the linear response to an external potential within the frame
the Bogolyubov model was already discussed in the literature in some
detail\cite{Pitaevskii} we want to reopen the discussion here, since
important features due to the self-consistency have been ignored until
now. The modifications shed a new light on the problem of supra-fluidity
of the Bogolyubov model. According to Hohenberg and Martin \cite{Hohenberg}
a sufficient and necessary condition for super-fluidity is that the
density response to an external potential in Fourier-space diverges
at vanishing frequency and wave vector. The Hugenholtz-Pines theorem
\cite{Hugenholtz-Pines} ensures this feature in the frame of the
exact hamilton theory. In other words, the existence of gap-less quasi-particles
in equilibrium is the key for the understanding of super-fluidity.
Of course the nature of the singularity may be complicated by the
fact, that multi-quasiparticle states have also a vanishing threshold. 

The problem is however much more complicated in the case of self-consistent
theories, like the HF or the Bogolyubov model, since s.c. parameters
(averages) are present in the s.c. Hamiltonian and they are themselves
modified by the external perturbation. In this sense one speaks about
an induced perturbation. Our next purpose is to analyze this response
in the case of Bogolyubov's model. We ignore here any dissipation,
therefore any statement about singularities of the response theory
has to be regarded with some caution. The inverse statement is however
true: if without dissipation one gets no supra-fluid properties, no
such properties will arise due to dissipation either.

\subsection{General formalism}

We consider now the time-dependent self-consistent linear response
to a time dependent potential $U(x,t)$ starting from the s.c. HF
equilibrium in the absence of the external potential. 

We have to take into account however the peculiarities of the {}``equilibrium''
distribution in the presence of spontaneous symmetry breaking. The
broken symmetry is the particle number conservation and therefore,
under condensation conditions, $H_{0}^{HF}$ does not commute with
the particle number operator $N$. Therefore the macrocanonical density
matrix $\frac{1}{Z}e^{-\beta(H_{HF}^{0}-\mu N)}$is not a stationary
state of the system in the usual sense \cite{Pitaevskii,Oli}. More
precisely, stationarity is reached by the {}``rotated'' density
matrix:

\[
\tilde{\rho}(t)\equiv e^{\frac{\imath}{\hbar}\mu Nt}\rho(t)e^{-\frac{\imath}{\hbar}\mu Nt}\]
 which satisfies the Liouville equation

\[
\imath\hbar\frac{\partial}{\partial t}\tilde{\rho}=[{\mathcal{H}}_{HF}(t),\tilde{\rho}]\]

generated by the {}``Hamiltonian''

\[
{\mathcal{H}}_{HF}(t)\equiv e^{\frac{\imath}{\hbar}\mu Nt}H_{HF}(t)e^{-\frac{\imath}{\hbar}\mu Nt}-\mu N\,.\]

Indeed, in the spirit of the Van Hove limit of weak coupling to a
thermal bath, $\tilde{\rho}(t)$ approaches, for large times, the
value $\frac{1}{Z}e^{-\beta(H_{HF}^{0}-\mu N)}$. In the absence of
the spontaneous symmetry breaking, the rotated density matrix is,
of course, identical to the density matrix itself.

We denote

\[
\langle\langle A\rangle\rangle_{t}\equiv Tr\left(\tilde{\rho}(t)A\right)\]
and \[
\langle A\rangle_{t}\equiv Tr\left(\rho(t)A\right)=e^{-\frac{\imath}{\hbar}m\mu t}\langle\langle A\rangle\rangle_{t};\;(m=0,1,2,...)\quad.\]

We assumed here that the operator $A$ has the structure $(\psi^{+})^{\bar{n}}\psi^{n}$.
Then $m=n-\bar{n}$ . Obviously the normal entities ($n=\bar{n}$)
are stationary in the peculiar {}``equilibrium'' state, however
the anomalous ones ($n\neq\bar{n}$) oscillate with a multiple of
the chemical potential. 

In the absence of the potential $U(x,t)$ the system is supposed to
be in {}``equilibrium'' \[
\tilde{\rho}|_{t=-\infty}=\frac{1}{Z}e^{-\beta{\mathcal{H}}_{HF}^{0}}\]

with translational invariant averages

\begin{eqnarray*}
\langle\langle\psi(x)\rangle\rangle^{0}\equiv P;\:\langle\langle\psi(x)^{+}\psi(x')\rangle\rangle^{0}\equiv\frac{1}{V}\sum_{k\neq0}e^{\imath k(x'-x)}f_{k}+|P|^{2};\\
\langle\langle\psi(x)\psi(x')\rangle\rangle^{0}\equiv(\frac{1}{V}\sum_{k\neq0}e^{\imath k(x+x')}F_{k}+P^{2})\end{eqnarray*}
 where ${\mathcal{F}}_{k}\equiv\frac{1}{V}\sum_{k\neq0}e^{\imath k(x+x')}F_{k}$
and $P^{2}$ have the same phase up to a multiple of $\frac{\pi}{2}$
and actually both may be taken to be real. Further it is assumed,
that the self-consistency relations in equilibrium are satisfied.

Again, in order to simplify the calculations, we treat explicitly
only the case of a contact particle interaction. In this case

\begin{eqnarray*}
{\mathcal{H}}_{HF}(t) & = & \int dx\left\{ \psi(x)^{+}\left(-\frac{\hbar^{2}}{2m}\nabla^{2}+U(x,t)-\mu\right)\psi(x)\right.\\
 & + & \frac{w}{2}\left[4\psi(x)^{+}\psi(x)\langle\psi(x)^{+}\psi(x)\rangle\right.\\
 & + & \left(\psi(x)\psi(x)\langle\psi(x)\psi(x)\rangle^{*}+h.c.\right)\\
 & - & \left.\left.\left(4\psi(x)\langle\psi(x)\rangle^{*}|\langle\psi(x)\rangle|^{2}+h.c.\right)\right]\right\} \,.\end{eqnarray*}

According to linear response theory, for any observable $A$ one has

\[
\delta\langle\langle A\rangle\rangle_{t}=\frac{1}{\imath\hbar}\int_{-\infty}^{t}dt'\langle\langle[A_{{\mathcal{H}}_{HF}^{0}}(t-t'),\delta{\mathcal{H}}_{HF}(t')]\rangle\rangle^{0}\]
 where \[
A_{{\mathcal{H}}_{HF}^{0}}(t)\equiv e^{\frac{\imath}{\hbar}{\mathcal{H}}_{HF}^{0}t}Ae^{-\frac{\imath}{\hbar}{\mathcal{H}}_{HF}^{0}t}\,,\]
 or in Fourier transforms\[
\delta\langle\langle A\rangle\rangle_{\omega}=\frac{1}{\imath\hbar}\int_{0}^{\infty}dte^{\imath\omega t}\langle\langle[A_{{\mathcal{H}}_{HF}^{0}}(t),\delta{\mathcal{H}}_{HF}(\omega)]\rangle\rangle^{0}\,.\]

Here $\delta{\mathcal{H}}_{HF}(t)$ contains all the terms of first
order (direct and induced) of the difference ${\mathcal{H}}_{HF}(t)-{\mathcal{H}}_{HF}^{0}$
. Thus, taking also into account that (with real $P$ ) \[
\delta\left(|\langle\langle\psi(x)\rangle\rangle_{t}|^{2}\langle\langle\psi(x)\rangle\rangle\right)=2P^{2}\delta\langle\langle\psi(x)\rangle\rangle+|P^{2}|\delta\langle\langle\psi(x)\rangle\rangle^{*}=P^{2}(2\delta\langle\langle\psi(x)\rangle\rangle+\delta\langle\langle\psi(x)\rangle\rangle^{*})\]
 we have

\begin{eqnarray*}
\delta{\mathcal{H}}_{HF}(t) & = & \int dx\left\{ \left(U(x,t)+2w\delta\langle\langle\psi(x)^{+}\psi(x)\rangle\rangle_{t}\right)\psi(x)^{+}\psi(x)\right.\\
 & + & \frac{w}{2}\left(\delta\langle\langle\psi(x)^{2}\rangle\rangle_{t}^{*}\psi(x)^{2}+h.c.\right)\\
 & - & \left.2wP^{2}\left[\left(\delta\langle\langle\psi(x)\rangle\rangle_{t}+2\delta\langle\langle\psi(x)\rangle\rangle_{t}^{*}\right)\psi(x)+h.c.\right]\right\} \,.\end{eqnarray*}

In what follows we restrict further the discussion to the Bogolyubov
version of the theory, where no anomalous correlations are taken into
account. Then we have

\begin{eqnarray*}
\delta{\mathcal{H}}_{HF}^{B}(t) & = & \int dx\left\{ \left(U(x,t)+2w\delta\langle\langle\psi(x)^{+}\psi(x)\rangle\rangle_{t}\right)\psi(x)^{+}\psi(x)\right.\\
 & + & \frac{w}{2}\left(2P\delta\langle\langle\psi(x)\rangle\rangle_{t}^{*}\psi(x)^{2}+h.c.\right)\\
 & - & \left.2wP^{2}\left[\left(\delta\langle\langle\psi(x)\rangle\rangle_{t}+2\delta\langle\langle\psi(x)\rangle\rangle_{t}^{*}\right)\psi(x)+h.c.\right]\right\} \;.\end{eqnarray*}

In order to simplify the formulae let us define now

\[
A=\left(\begin{array}{c}
\psi(x)^{+}\psi(x)\\
P\psi(x)\\
P\psi(x)^{+}\end{array}\right)\,,\qquad B=\left(\begin{array}{c}
\psi(x)^{+}\psi(x)\\
\frac{1}{2}\psi(x)^{2}-3P\psi(x)\\
\frac{1}{2}\psi(x)^{+2}-3P\psi(x)^{+}\end{array}\right)\]

and

\[
\delta{\mathcal{H}}_{HF}^{B}(t)=\sum_{\alpha}2w\int dxB_{\alpha}(x)^{+}\delta\langle\langle A_{\alpha}(x)\rangle\rangle+\int dxU(x,t)B_{1}(x)\,.\]

Then, after Fourier transforming ($\tilde{f}(k,\omega)=\int dx\int dte^{-\imath(kx-\omega t)}f(x,t)$
)

\[
\delta\langle\langle A_{\alpha}\rangle\rangle_{k,\omega}=\sum_{\alpha'}\chi_{\alpha\alpha'}(k,\omega)\delta\langle\langle A_{\alpha'}\rangle\rangle_{k,\omega}+\chi_{\alpha1}(k,\omega)\frac{1}{2w}\tilde{U}(k,\omega)\]

with \[
\chi_{\alpha\alpha'}(k,\omega)=\frac{2w}{\imath\hbar}\int_{0}^{\infty}dt\int dxe^{-\imath(kx-\omega t)}\langle\langle[A_{\alpha{\mathcal{H}}_{HF}^{0}}(x,t),B_{\alpha'}^{+}(0,0)]\rangle\rangle^{eq}\,.\]

Remark: By the interpretation of the susceptibility one has to perform
eventually also a shift with $\frac{\mu}{\hbar}m$ $(m=0,1,2,...)in$
the case of particle number non-conserving terms!

The relation of the averages of the different entities $A_{\alpha}$
to the external potential $\tilde{U}(k,\omega)$, with matrix notations,
is given by

\[
\delta\langle\langle A_{\alpha}\rangle\rangle_{k,\omega}=\left(\frac{\chi(k,\omega)}{1-\chi(k,\omega)}\right)_{\alpha1}\frac{1}{2w}\tilde{U}(k,\omega)=\Xi_{\alpha1}(k,\omega)\frac{1}{2w}\tilde{U}(k,\omega)\:.\]

\subsection{The structure of \protect$\chi(k,\omega)$ at \protect$T=0^{\circ}$.}

With the notation $P^{2}=\xi n$ ($0<\xi<1$ ) one gets $\mu=(2-\xi)nw$.
We chose $2nw$ as natural unit for energy and $\kappa\equiv\sqrt{\frac{2mnw}{\hbar^{2}}}$
for momenta. The new variables are :\[
\mu\to1-\frac{1}{2}\xi\]

\[
e_{q}\to q^{2}+1\]

\[
\epsilon_{q}\to\sqrt{\xi q^{2}+q^{4}}\]

\[
u_{q}=\sqrt{\frac{1}{2}\left(\frac{q^{2}+\frac{1}{2}\xi}{\sqrt{\xi q^{2}+q^{4}}}+1\right);\:}v_{q}=-\sqrt{\frac{1}{2}\left(\frac{q^{2}+\frac{1}{2}\xi}{\sqrt{\xi q^{2}+q^{4}}}-1\right)}\]

\[
\hbar\omega\to\bar{\omega}=\frac{\hbar\omega}{2nw}\,.\]

Since at $T\to0^{\circ}$ we have

\[
\int d\vec{p}\frac{e_{p}-\mu}{\epsilon_{p}}\frac{1}{(e^{\beta\epsilon_{p}}-1)}\to0\]
 the self-consistency equation looks at $T=0^{\circ}$ in the new
units as as:

\[
\xi+\frac{1}{(2\pi)^{3}}\frac{\kappa^{3}}{n}\int d^{3}\vec{q}\frac{1}{2}(\frac{q^{2}+\frac{1}{2}\xi}{\epsilon_{q}}-1)=1\,\]

giving rise to the simple dependence between parameter $\xi$ and
the density $n$

\begin{equation}
\frac{1}{(2\pi)^{2}}\frac{\kappa^{3}}{n}=\frac{6(1-\xi)}{\xi^{\frac{3}{2}}}\,.\label{xi-n}\end{equation}

We have to apply the commutations rules:

\[
[\alpha_{q},\alpha_{p}]=0;\:[\alpha_{q},\alpha_{p}^{+}]=\delta_{q,p}\]

\[
[\alpha_{q,}\alpha_{p}^{+}\alpha_{p'}^{+}]=\alpha_{q}(\delta_{q,p}+\delta_{q,p'})\]

\[
[\alpha_{q}\alpha_{q'},\alpha_{p}\alpha_{p'}]=0;\:[\alpha_{q}\alpha_{q'},\alpha_{p}^{+}\alpha_{p'}^{+}]=\delta_{p,q'}\alpha_{q}\alpha_{p'}^{+}+\delta_{q',p'}\alpha_{q}\alpha_{p}^{+}+\delta_{q,p}\alpha_{p'}^{+}\alpha_{q'}+\delta_{q,p'}\alpha_{p}^{+}\alpha_{q'}\]
 and since at $T=0^{\circ}$

\[
\langle0|\alpha_{q}|0\rangle=\langle0|\alpha_{q}^{+}\alpha_{q'}|0\rangle=0\]

the only non-vanishing terms are\[
\langle0|[\alpha_{q},\alpha_{p}^{+}]|0\rangle=\delta_{q,p};\:\langle0|[\alpha_{q}\alpha_{q'},\alpha_{p}^{+}\alpha_{p'}^{+}]|0\rangle=\delta_{p,q}\delta_{p',q'}+\delta_{p,q'}\delta_{p',q}\]

and their conjugates.

Further \[
\tilde{A}(k,t)=\]

\[
\left(\begin{array}{c}
\!\sum_{q}(v_{q-k}^{*}u_{q}\alpha_{k-q}\alpha_{q}e^{-\imath(\epsilon_{k-q}\!+\!\epsilon_{q})t}+v_{q+k}u_{q}^{*}\alpha_{q}^{+}\alpha_{-k-q}^{+}e^{\imath(\epsilon_{k+q}\!+\!\epsilon_{q})t})\!+\!\sqrt{V}P(u_{k}\alpha_{k}e^{-\imath\epsilon_{k}t}\!+\! v_{k}\alpha_{-k}^{+}e^{\imath\epsilon_{k}t}+h.c.(-k))\!\!\\
P\sqrt{V}(u_{k}\alpha_{k}e^{-\imath\epsilon_{k}t}+v_{k}\alpha_{-k}^{+}e^{\imath\epsilon_{k}t})\\
P\sqrt{V}(u_{k}^{*}\alpha_{k}^{+}e^{\imath\epsilon_{k}t}+v_{k}^{*}\alpha_{-k}e^{-\imath\epsilon_{k}t})\end{array}\right)\]

\[
B(0)^{+}=\left(\begin{array}{c}
\psi(0)^{+}\psi(0)\\
\frac{1}{2}\psi(0)^{+2}-3P\psi(0)^{+}\\
\frac{1}{2}\psi(0)^{2}-3P\psi(x)\end{array}\right)=\left(\begin{array}{c}
\frac{1}{V}\sum_{p,p'}\hat{b}_{p}^{+}\hat{b}_{p'}+\frac{P}{\sqrt{V}}\sum_{p}(\hat{b}_{p}+\hat{b}_{p}^{+})\\
\frac{1}{2V}\sum_{p,p'}\hat{b}_{p'}^{+}\hat{b}_{p}^{+}+\frac{P}{\sqrt{V}}\sum_{p}\hat{b}_{p}^{+}-3\frac{P}{\sqrt{V}}\sum_{p}\hat{b}_{p}^{+}\\
\frac{1}{2}\frac{1}{V}\sum_{p,p'}\hat{b}_{p'}\hat{b}_{p}+\frac{P}{\sqrt{V}}\sum_{p}\hat{b}_{p}-3\frac{P}{\sqrt{V}}\sum_{p}\hat{b}_{p}\end{array}\right)\]

or

\[
B(0)^{+}=\left(\begin{array}{c}
\frac{1}{V}\sum_{p,p'}(v_{p}^{*}u_{p'}\alpha_{-p}\alpha_{p'}+u_{p}^{*}v_{p'}\alpha_{p}^{+}\alpha_{-p'}^{+})+\frac{P}{\sqrt{V}}\sum_{p}(u_{p}\alpha_{p}+v_{p}\alpha_{-p}^{+}+h.c.)\\
\frac{1}{2V}\sum_{p,p'}(u_{p}^{*}u_{p'}^{*}\alpha_{p'}^{+}\alpha_{p}^{+}+v_{-p}^{*}v_{-p'}^{*}\alpha_{p'}\alpha_{p}+u_{p}^{*}v_{-p'}^{*}\alpha_{p}^{+}\alpha_{p'})-2\frac{P}{\sqrt{V}}\sum_{p}(u_{p}^{*}\alpha_{p}^{+}+v_{p}^{*}\alpha_{-p})\\
\frac{1}{2V}\sum_{p,p'}(u_{p}u_{p'}\alpha_{p'}\alpha_{p}+v_{-p}v_{-p'}\alpha_{p'}^{+}\alpha_{p}^{+}+u_{p}v_{-p'}\alpha_{p'}^{+}\alpha_{p})-2\frac{P}{\sqrt{V}}\sum_{p}(u_{p}\alpha_{p}+v_{p}\alpha_{-p}^{+})\end{array}\right)\,.\]

At zero temperature $T=0^{\circ}$ we have to average over the ground
state (vacuum of quasi-particles)

\[
\chi_{\alpha\alpha'}(k,t)\sim\langle0|[A_{\alpha{\mathcal{H}}_{HF}^{eq}}(x,t),B_{\alpha'}^{+}(0,0)]|0\rangle^{eq}\]

\[
\chi_{11}(k,t)\sim P^{2}e^{-\imath\epsilon_{k}t}|u_{k}+v_{-k}^{*}|^{2}+\frac{1}{V}\sum_{q}e^{-\imath(\epsilon_{k-q}+\epsilon_{q})t}u_{q}v_{q-k}^{*}(u_{k-q}^{*}v_{q}+v_{k-q}u_{q}^{*})-c.c(-k)\]

\begin{eqnarray*}
\chi_{12}(k,t)\sim\frac{1}{V}\sum_{q}\left(e^{-\imath(\epsilon_{k-q}+\epsilon_{q})t}v_{q-k}^{*}|u_{q}|^{2}u_{k-q}^{*}-e^{\imath(\epsilon_{k+q}+\epsilon_{q})t}u_{q}^{*}|v_{q+k}|^{2}v_{-q}^{*}\right)\\
-2P^{2}\left((u_{k}+v_{-k}^{*})u_{k}^{*}e^{-\imath\epsilon_{k}t}-(u_{-k}^{*}+v_{k})v_{k}^{*}e^{\imath\epsilon_{k}t}\right)\end{eqnarray*}

\begin{eqnarray*}
\chi_{13}(k,t)\sim\frac{-1}{V}\sum_{q}\left(e^{\imath(\epsilon_{k+q}+\epsilon_{q})t}|u_{q}|^{2}v_{q+k}u_{-q-k}-e^{-\imath(\epsilon_{k-q}+\epsilon_{q})t}|v_{q-k}|^{2}u_{q}v_{-q}\right)\\
-2P^{2}\left((u_{k}+v_{-k}^{*})v_{-k}e^{-\imath\epsilon_{k}t}-(u_{-k}^{*}+v_{k})u_{-k}e^{\imath\epsilon_{k}t}\right)\end{eqnarray*}

\[
\chi_{21}(k,t)\sim P^{2}\left(u_{k}(v_{-k}+u_{k}^{*})e^{-\imath\epsilon_{k}t}-v_{k}(u_{-k}+v_{k}^{*})e^{\imath\epsilon_{k}t}\right)\]

\[
\chi_{22}(k,t)\sim2P^{2}\left(|v_{k}|^{2}e^{\imath\epsilon_{k}t}-|u_{k}|^{2}e^{-\imath\epsilon_{k}t}\right)\]

\[
\chi_{23}(k,t)\sim2P^{2}\left(v_{k}u_{-k}e^{\imath\epsilon_{k}t}-u_{k}v_{-k}e^{-\imath\epsilon_{k}t}\right)\]

\[
\chi_{31}(k,t)\sim P^{2}\left(v_{k}^{*}(v_{k}+u_{-k}^{*})e^{-\imath\epsilon_{k}t}-u_{k}^{*}(u_{k}+v_{-k}^{*})e^{\imath\epsilon_{k}t}\right)\]

\[
\chi_{32}(k,t)\sim2P^{2}\left(u_{k}^{*}v_{-k}^{*}e^{\imath\epsilon_{k}t}-v_{k}^{*}u_{-k}^{*}e^{-\imath\epsilon_{k}t}\right)\]

\[
\chi_{33}(k,t)\sim2P^{2}\left(|u_{k}|^{2}e^{\imath\epsilon_{k}t}-|v_{k}|^{2}e^{-\imath\epsilon_{k}t}\right)\]

and

\begin{eqnarray*}
\chi_{11}(k,\bar{\omega})=\xi|u_{k}+v_{k}|^{2}\left(\frac{1}{\bar{\omega}+\epsilon_{k}}-\frac{1}{\bar{\omega}-\epsilon_{k}}\right)\\
+\frac{\kappa^{3}}{n(2\pi)^{3}}\int d\vec{q}\left[\frac{1}{\bar{\omega}+\epsilon_{q}+\epsilon_{q-k}}-\frac{1}{\bar{\omega}-\epsilon_{q}-\epsilon_{q-k}}\right]u_{q}v_{q-k}(u_{q-k}v_{q}+v_{q-k}u_{q})\end{eqnarray*}

\begin{eqnarray*}
\chi_{12}(k,\bar{\omega})=2\xi(u_{k}+v_{k})\left(u_{k}\frac{1}{\bar{\omega}-\epsilon_{k}}-v_{k}\frac{1}{\bar{\omega}+\epsilon_{k}}\right)\\
-\frac{\kappa^{3}}{n(2\pi)^{3}}\int d\vec{q}\left(\frac{1}{\bar{\omega}-\epsilon_{q}-\epsilon_{q-k}}v_{q-k}u_{q}^{2}u_{q-k}-\frac{1}{\bar{\omega}+\epsilon_{q}+\epsilon_{q-k}}u_{q}v_{q-k}^{2}v_{q}\right)\end{eqnarray*}

\begin{eqnarray*}
\chi_{13}(k,\bar{\omega})=2\xi(u_{k}+v_{k})\left(v_{k}\frac{1}{\bar{\omega}-\epsilon_{k}}-u_{k}\frac{1}{\bar{\omega}+\epsilon_{k}}\right)\\
-\frac{\kappa^{3}}{n(2\pi)^{3}}\int d\vec{q}\left(u_{q}v_{q-k}^{2}v_{q}\frac{1}{\bar{\omega}-\epsilon_{q}-\epsilon_{q-k}}-\frac{1}{\bar{\omega}+\epsilon_{q}+\epsilon_{q-k}}v_{q-k}u_{q}^{2}u_{q-k}\right)\end{eqnarray*}

\[
\chi_{21}(k,\bar{\omega})=-\xi(u_{k}+v_{k})\left(u_{k}\frac{1}{\bar{\omega}-\epsilon_{k}}-v_{k}\frac{1}{\bar{\omega}+\epsilon_{k}}\right)\]

\[
\chi_{22}(k,\bar{\omega})=-2\xi\left(|v_{k}|^{2}\frac{1}{\bar{\omega}+\epsilon_{k}}-|u_{k}|^{2}\frac{1}{\bar{\omega}-\epsilon_{k}}\right)\]

\[
\chi_{23}(k,\bar{\omega})=-2\xi u_{k}v_{k}\left(\frac{1}{\bar{\omega}+\epsilon_{k}}-\frac{1}{\bar{\omega}-\epsilon_{k}}\right)\]

\[
\chi_{31}(k,\bar{\omega})=-\xi(u_{k}+v_{k})\left(v_{k}\frac{1}{\bar{\omega}-\epsilon_{k}}-u_{k}\frac{1}{\bar{\omega}+\epsilon_{k}}\right)\]

\[
\chi_{32}(k,\bar{\omega})=-2\xi u_{k}v_{k}\left(\frac{1}{\bar{\omega}+\epsilon_{k}}-\frac{1}{\bar{\omega}-\epsilon_{k}}\right)\]

\[
\chi_{33}(k,\bar{\omega})=-2\xi\left(|u_{k}|^{2}\frac{1}{\bar{\omega}+\epsilon_{k}}-|v_{k}|^{2}\frac{1}{\bar{\omega}-\epsilon_{k}}\right)\,.\]

As it may be seen, the susceptibilities have poles in the complex
$\bar{\omega}$ plane at $\epsilon_{k}$ and cuts on the real axis
from $2\epsilon_{k/2}$ to $\infty$ , respectively from $-2\epsilon_{k/2}$
to -$\infty$ . Since $\epsilon_{k}\geq2\epsilon_{k/2}$, the poles
are embedded in the cuts.

The density-density correlation $\chi_{11}(k,\omega)$ of the Bogolyubov
model was already discussed at arbitrary temperatures in Ref.\cite{Pitaevskii}.
Our accent is on taking into account the induced perturbation and
therefore needs the knowledge of the whole matrix $\chi(k,\omega)$.

\subsection{The density-density response at $\vec{k}=0$ and $\omega\to0$.}

At $\vec{k}=0$, with the notations

\[
c_{1}(\bar{\omega})\equiv\frac{2\kappa^{3}}{n(2\pi)^{3}}\int d\vec{q}\left[\frac{1}{\bar{\omega}+2\epsilon_{q}}-\frac{1}{\bar{\omega}-2\epsilon_{q}}\right]u_{q}^{2}v_{q}^{2}\]

\[
c_{2}(\bar{\omega})\equiv\frac{\kappa^{3}}{n(2\pi)^{3}}\int d\vec{q}\left(\frac{1}{\bar{\omega}+2\epsilon_{q}}u_{q}v_{q}^{3}-\frac{1}{\bar{\omega}-2\epsilon_{q}}v_{q}u_{q}^{3}\right)\]

\[
c_{2}(\bar{\omega})\equiv\frac{\kappa^{3}}{n(2\pi)^{3}}\int d\vec{q}\left(\frac{1}{\bar{\omega}+2\epsilon_{q}}u_{q}^{3}v_{q}-\frac{1}{\bar{\omega}-2\epsilon_{q}}v_{q}^{3}u_{q}\right)\]

one gets \begin{equation}
\chi(\omega)=\left(\begin{array}{ccc}
c_{1}(\bar{\omega}) & \frac{2\xi}{\bar{\omega}}+c_{2}(\bar{\omega}) & -\frac{2\xi}{\bar{\omega}}+c_{3}(\bar{\omega})\\
-\frac{\xi}{\bar{\omega}} & \frac{2\xi}{\bar{\omega}}+\frac{\xi^{2}}{\bar{\omega}^{2}} & -\frac{\xi^{2}}{\bar{\omega}^{2}}\\
\frac{\xi}{\bar{\omega}} & -\frac{\xi^{2}}{\bar{\omega}^{2}} & -\frac{2\xi}{\bar{\omega}}+\frac{\xi^{2}}{\bar{\omega}^{2}}\end{array}\right)\,.\label{chitab}\end{equation}

Of course the coefficients of the integrals are not independent entities
and are related to the parameter $\xi$ by the self-consistency relation
Eq.(\ref{xi-n}).

Remark, that \[
\chi_{11}(0,\omega)\sim\log{\omega}\qquad\text{for}\qquad\omega\to0\]
 since the residua of its poles vanish at $k=0$. Thus the naive response
theory\cite{Pitaevskii}, which ignores subtleties of the induced
perturbation leads indeed to a singular response, although the singularity
is only a logarithmic one. The cuts responsible for this behavior
are due to excitations of two quasi-particles. These are however not
present in the near-equilibrium oscillations of the solution of the
Gross-Pitaevski equation. 

However, taking into account the induced perturbation from Eq.(\ref{chitab})
it follows \begin{equation}
\Xi_{11}(0,\omega)=\left(\frac{\chi(0,\omega)}{1-\chi(0,\omega)}\right)_{11}=\frac{-2\xi^{2}(2+3c_{1}(\omega)+c_{2}(\omega)+c_{3}(\omega))+\xi\omega(c_{3}(\omega)-c_{2}(\omega))+\omega^{2}c_{1}(\omega)}{2\xi^{2}(-1+3c_{1}(\omega)+c_{2}(\omega)+c_{3}(\omega))-\xi\omega(c_{3}(\omega)-c_{2}(\omega))+\omega^{2}(1-c_{1}(\omega))}\,.\label{dens-dens}\end{equation}

Since $c_{1}(\omega),\quad c_{2}(\omega),\quad c_{3}(\omega)$ behave
like $\log{\omega}$ for $\omega\to0$, it follows \begin{equation}
\lim_{\omega\to0}\Xi_{11}(0,\omega)=-1\,.\label{dens-dens0}\end{equation}

Thus we may conclude, that although the Bogolyubov model has quasi-particles
with vanishing energy at $q\to0$, the response to an external potential
is not at all singular! 

The special value $-1$ of the effective susceptibility implies that
the Fourier transform of the effective induced potential for the non-condensate
defined as $U_{eff}^{nc}(x,t)\equiv U(x,t)+2w\delta\langle\psi(x)^{+}\psi(x)\rangle_{t}$
vanishes in the limit $k\to0$ and $\omega\to0$. The factor $2$
in this definition stems from the fact, that in the case of a contact
interaction the direct and exchange terms are identical. On the other
hand, the effective induced potential for the condensate is different
$U_{eff}^{c}(x,t)\equiv U_{eff}^{nc}(x,t)-w\delta|\langle\psi(x)\rangle_{t}|^{2}$
and does not vanish in this limit.

Superfluidity would require diverging density response to a finite
external perturbation and this does not happen.

\section{Conclusions}

We have shown, that although the Bogolyubov model leads to a gap-less
quasiparticle model, if one takes into account the induced perturbation,
it leads to no low frequency, low wave-vector singularity of the response
to a perturbation by an external potential. In this sense it does
not fulfill the hopes to satisfy the criterion of Martin and Hohenberg
for superfluidity. Earlier treatments of the linear response in Bose
condensed systems either ignored the induced perturbation \cite{Pitaevskii}
or did not discuss in all detail his consequences, their attention
being concentrated on improving the Hartree-Fock theory \cite{Giorgini,Rusch-Burnett}.
As it is known, the Bogolyubov model is also not the best s.c. approximation
from the thermodynamical point of view, but still obeys the Huggenholz-Pines
theorem, which the thermodynamically better Hartree-Fock theory does
not obey. Seemingly none of the simple self-consistent approximations
is quite satisfactory and better models have to be developed. \\

One of the  authors (A.M.B) thanks the Deutsche Forschungsgemeinschaft for
the generous support allowing his stay at the Frankfurt University.

\end{document}